\documentstyle[12pt]{article}
\begin{document}
\large
\begin{center}
{\bf $\pi\pm \leftrightarrow K\pm$ Meson Vacuum Transitions
(Oscillations) in Diagram Approach in the Model of Dynamical Analogy of
the Cabibbo-Kobayashi-Maskawa Matrices}\\
\vspace{2cm}
Kh.M. Beshtoev\\
\vspace{2cm}
Joint Institute for Nuclear Research, Joliot Curie 6\\
141980 Dubna, Moscow region and Inst. Appl. Math. and Autom. \\
KBSRC of RAS, Nalchik, Kabardino-Balk. Rep., Russia\\
\vspace{2cm}
\end{center}
\par
{\bf Abstract}
\par
The elements of the theory of vacuum oscillations and the model of
dynamical expansion of the theory of weak interactions works at the
tree level, i.e. the model of dynamical analogy of
Cabibbo-Kobayashi- Maskawa matrices and its further development, are given.
It is shown that the quarks and massive vector bosons must be structural
and these structural particles (subparticles) must interact to
generate quark and vector boson masses.
In this case  the problem of singularity cancellations does not arise
in this model.
It is also shown
that, for self consistency of the theory, the weak decays of $K$ mesons
must go through massive vector boson $B$ but not $W$ boson.
\par
In the framework of this model the probability of $\pi \leftrightarrow K$
transitions (oscillations) in the diagram approach is computed.
These transitions are virtual since masses of $\pi$ and $K$ mesons
differ considerably.  These transitions (oscillations) can be registered
through $K$ decays after transitions of virtual $K$ mesons to their own
mass shell by using their quasielastic strong interactions.

\section{Introduction}

The vacuum oscillation of neutral $K$ mesons is well investigated at
the present time [1]. This oscillation is the result of $d, s$ quark
mixings and is described by Cabibbo-Kobayashi-Maskawa matrices [2].
The angle mixing $\theta$ of neutral $K$ mesons is $\theta = 45^O$ since
$K^o, \bar K^o$ masses are equal (see $CPT$ theorem). Besides, since
their masses are equal, these oscillations are real, i.e. their
transitions to each other go without suppression. Oscillations
of two particles having the masses overlapping their widths were discussed
in works [3]. Then we calculated probabilities of $\pi \leftrightarrow K$
oscillations in an approach where the phase volume of
particles at these transitions is taken into account [4,5].
\par
This work is devoted to the development of the model of dynamical analogy of
Cabibbo-Kobayashi- Maskawa matrices [6] and to the calculation of
probabilities of $\pi \leftrightarrow K$ oscillations in framework of
this model in the diagram approach [7] which was used while calculation of
$K^o \leftrightarrow \bar K^o$ oscillations.
\par
At first, we will consider the general elements of the theory of
oscillations, elements of the model of dynamical analogy of
Cabibbo-Kobayashi- Maskawa matrices and its further development, then
come to the calculation of probabilities of $\pi \leftrightarrow K$
transitions.
\par
As it is stressed in previous works [4,5] these transitions are virtual
since masses of $\pi$ and $K$ mesons differ considerably.
And we can make these virtual transitions real through their strong
interactions, i.e. bring them up on the own mass shell through strong
interactions after the weak interaction transforming $\pi$ mesons
in virtual $K$ mesons.
\par
Let us to consider the general elements of the theory of oscillations.

\section{Probabilities of Real and Virtual Vacuum
$\pi \leftrightarrow K$ Oscillations (Transitions)}
\par
The mass matrix of $\pi$ and $K$ mesons has the form
$$
\left(\begin{array}{cc} m_\pi & 0 \\ 0 & m_K  \end{array} \right) .
\eqno(1)
$$
\par
Due to the presence of strangeness violation in the weak interactions,
a nondiagonal term appears in this matrix and then this mass matrix is
transformed in the following nondiagonal matrix:
$$
\left(\begin{array}{cc}m_\pi & m_{\pi K} \\ m_{\pi K} & m_K  \end{array}
\right)   ,
\eqno(2)
$$
which is diagonalized by turning through the angle $\beta$ and then
$$
\left(\begin{array}{cc}m_\pi & m_{\pi K} \\ m_{\pi K} & m_K  \end{array}
\right)   \to
\left(\begin{array}{cc}m_1 & 0 \\ 0 & m_2  \end{array}
\right)
\eqno(3)
$$
where
$$
tg 2\beta = \frac{2m_{\pi K}}{\mid m_\pi - m_K \mid}   ,
$$
$$
sin 2\beta = \frac{2m_{\pi K}}{\sqrt{(m_\pi - m_K)^2 +(2m_{\pi K})^2}}  ,
\eqno(4)
$$
$$
m_{1,2} = \frac{1}{2}((m_{\pi} - m_{K}) \pm
\sqrt{(m_{\pi} - m_{K})^2 + 4(m_{\pi K})^2}) .
$$
\par
It is interesting to remark that expression (4) can be obtained from
the Breit-Wigner distribution [8]
$$
P \sim \frac{(\Gamma/2)^2}{(E - E_0)^2 + (\Gamma/2)^2}
\eqno(5)
$$
by using the following substitutions:
$$
E = m_K,\hspace{0.2cm} E_0 = m_\pi,\hspace{0.2cm} \Gamma/2 = 2m_{\pi K} ,
\eqno(6)
$$
where $\Gamma \equiv W(... )$.
\par
If the mass matrix contains masses in a squared form, then oscillations
(or mixings) will be described by the expressions (3)-(6) with the following
substitutions:
$$
m_\pi \to m_\pi^2, m_K \to m_K^2, m_{\pi K} \to m_{\pi K}^2
$$
\par
Here two cases of $\pi, K$ oscillations [4] take place: real and
virtual oscillations.
\par
1. If we consider the real transition of $\pi$ into $K$ mesons, then
$$
sin^2 2\beta \cong \frac{4m^2_{\pi K}}{(m_\pi - m_K)^2} \cong 0 ,
\eqno(7)
$$
i.e. the probability of the real transition of $\pi$ mesons into $K$
mesons through weak interactions is very small since $m_{\pi K}$ is
very small.
\par
How can we understand this real $\pi \rightarrow K$ transition?
\par

If $2m_{\pi K} = \frac{\Gamma}{2}$ is not zero, then it means that
the mean mass of $\pi$ meson is $m_\pi$ and this mass is distributed
by $sin^2 2\beta$ (or by the Breit-Wigner formula) and the probability
of the $\pi \rightarrow K$ transition differs from zero. So, this is
a solution of the problem of origin of mixing angle in the theory
of vacuum oscillation.
\par
In this case the probability of $\pi \rightarrow K$ transition (oscillation)
is described by the following expression:
$$
P(\pi \rightarrow K, t) =  sin^2 2\beta
sin^2 \left[\pi t \frac{m_K^2}{2 p} \right ] ,
\eqno(8)
$$
where $p$ is momentum of $\pi$ meson.
\par
2. If we consider the virtual transition of $\pi$ into $K$ meson, then,
since $m_\pi = m_K$,
$$
tg 2\beta = \infty  ,
$$
i.e. $\beta = \pi/4$, then
$$
sin^2 2\beta = 1     .
\eqno(9)
$$
\par
In this case the probability of $\pi \rightarrow K$ transition (oscillation)
is described by the following expression:
$$
P(\pi \rightarrow K, t) =
sin^2 \left[\pi \frac{L}{L_{osc}} \right ] ,
\eqno(10)
$$
where $L = v t$, $v$- is a velocity of $\pi$ meson,
at $v \cong c$ $L \cong c t$,
$$
L_{osc} = {2.48 p_{\pi}(MeV) \over \mid m_1^2 - m_2^2 \mid (eV^2)} m .
$$
\par
Let us consider elements of the model of dynamical analogy of
Cabibbo-Kobayashi-Maskawa matrices and its development.

\section {Elements of
the Model of Dynamical Analogy of the Cabibbo-Kobayashi-Maskawa
Matrices and Its Development}

In the case of three  families of quarks, the current $J ^{\mu }$ has
the following form:
\par
$$ J^{\mu } = ( \bar{u}  \bar{c}  \bar{t} )_{L} \gamma ^{\mu} V \left
( \begin{array}{c}
d\\ s\\ b\\ \end{array} \right)_{L}
\eqno(11)
$$
\begin{displaymath} {V =
\left( \begin{array}{ccc} V_{ud}& V_{us}& V_{ub}\\ V_{cd}& V_{cs}& V_{cb}\\
V_{td}& V_{ts}& V_{tb} \end{array} \right)}  ,
\end{displaymath}

\noindent
where $V$ is Kobayashi-Maskawa  matrix [2].
\par
Mixings of the $d, s, b$ quarks are not connected with the weak
interaction (i.e., with $W^\pm, Z^o$ bosons exchanges). From equation (1)
it is well seen that mixings of the $d, s,b$ quarks and exchange of
$W^\pm, Z^o$ bosons take place in an independent manner (i.e.,
if matrix $V$ were diagonal, mixings of the $d, s, b$ quarks would not
have taken place).
\par
If the mechanism of this mixings is realized independently of the weak
interaction ($W^\pm, Z^o$ boson exchange) with a probability determined
by the mixing angles $\theta, \beta, \gamma, \delta$ (see below), then
this violation could be found in the strong and electromagnetic
interactions of the quarks as a clear violations of isospin,
strangeness and beauty. But, the available experimental results have
shown,
that there is no clear violations of the number conservations in strong
and electromagnetic interactions of the quarks. Then we must connect the
non-conservation of isospins, strangeness and beauty (or mixings of
the $d, s, b$ quarks) with some type of interaction mixings of the quarks.
We can do it introducing (together with the $W^\pm, Z^o$ bosons) the
heavier vector bosons $B^\pm, C^\pm, D^\pm, E^\pm$ which interact with
the $d, s, b$ quarks with violation of isospin, strangeness and beauty.
\par
We shall choose  parametrization of  matrix $V$  in the form offered by
Maiani [9]
\par
\begin{displaymath}{V = \left( \begin{array} {ccc}1& 0 & 0 \\
0 & c_{\gamma} & s_{\gamma} \\ 0 & -s_{\gamma} & c_{\gamma} \\
\end{array} \right) \left( \begin{array}{ccc} c_{\beta} & 0 &
s_{\beta} \exp(-i\delta) \\ 0 & 1 & 0 \\ -s_{\beta} \exp(i\delta) &
0 & c_{\beta} \end{array} \right) \left( \begin{array}{ccc} c_{\theta}
& s_{\theta} & 0 \\ -s_{\theta} & c_{\theta} & 0 \\ 0 & 0 & 1
\end{array}\right)} , \end{displaymath}
\par
$$
c_{\theta} = \cos {\theta } , s_{\theta} =\sin{\theta} , \exp(i\delta)
= \cos{\delta } + i \sin{\delta} .
\eqno(12)
$$
\noindent
To the nondiagonal terms in (12), which are  responsible for  mixing
of the $d, s, b$- quarks and $CP$-violation in the three matrices, we
shall make correspond four doublets of vector bosons $B^{\pm},
C^{\pm}, D^{\pm}, E^{\pm} $ whose contributions are parametrized  by
four  angles $\theta ,\beta ,\gamma ,\delta $ . It is supposed that
the  real part of $Re(s_{\beta} \exp(i\delta)) = s_\beta \cos{\delta} $
corresponds to the vector boson $C^{\pm}$ , and  the imaginary part of
$Im(s_{\beta } \exp(i\delta)) = s_{\beta}\sin \delta$ corresponds to
the vector boson $E^{\pm}$ (the couple constant  of $E$  is an imaginary
value !). Then, when $q^{2}<< m^{2}_{W}$ , we get:
\par
$$
\tan{\theta } \cong \frac{m^{2}_{W} g^{2}_{B}}{ m^{2}_{B} g^{2}_{W}},
\quad
\tan{\beta} \cong \frac{m^{2}_{W} g^{2}_{C}}{ m^{2}_{C} g^{2}_{W}} ,
$$
$$
\tan{\gamma} \cong \frac{m^{2}_{W} g^{2}_{D}}{ m^{2}_{D} g^{2}_{W}} ,
\quad
\tan{\delta} \cong \frac{m^{2}_{W} g^{2}_{E}}{m^{2}_{E} g^{2}_{W}} .
\eqno(13)
$$
\par
\noindent
If $g_{B^{\pm}} \cong  g_{C^{\pm}} \cong g_{D^{\pm}} \cong  g_{E ^{\pm}}
\cong g_{W^{\pm}}$ , then
$$ \tan{\theta } \cong \frac{ m^{2}_{W}}{ m^{2}_{B}} ,
\quad \tan{\beta} \cong \frac{m^{2}_{W}}{ m^{2}_{C}}  , $$

$$\tan{\gamma} \cong \frac{m^{2}_{W}}{ m^{2}_{D}}  ,
\quad
\tan{\delta} \cong \frac{m^{2}_{W}}{ m^{2}_{E}}  .
\eqno(14)
$$
\par
\medskip
Concerning the  neutral  vector  bosons $B^{0} , C^{0} , D^{0}, E^{0}$,
the neutral scalar bosons $B^{'0},C^{'0},D^{'0} ,E^{'0}$ and the
GIM mechanism [10], can repeat the same arguments given
in the previous work [6].
\par
The proposed Lagrangian for expansion of the weak interaction theory
(without CP-violation) has the following form:
$$
L_{int} = i
\sum_{i} g_{i} (J^{i,\alpha}A^{i}_{\alpha} + c. c. ) ,
\eqno(15)
$$
where
$J^{i,\alpha} = \bar\psi_{i,L} \gamma^{\alpha} T \varphi_{i,L} $,
\begin{displaymath}{T = \left(\begin{array}{cc}  0 & 1\\ -1
& 0  \end{array} \right)} , \end{displaymath}

\hspace{5.3cm}$ i = 1 \hspace{0.5cm} i = 2 \hspace{0.5cm} i = 3$
\begin{displaymath}{\psi_{i,L} =
\left( \begin{array}{c} u \\ c \end{array} \right)_{L} ,
\left( \begin{array}{c} u \\ t \end{array} \right)_{L},
\left( \begin{array}{c} c \\ t \end{array} \right)_{L}},
\end{displaymath}
\hspace{5.73cm}$ i = 1 \hspace{0.5cm} i = 2 \hspace{0.5cm} i = 3$
\begin{displaymath}{\varphi_{i,L} =
\left( \begin{array}{c} d \\ s \end{array} \right)_{L} ,
\left( \begin{array}{c} d \\ b \end{array} \right)_{L},
\left( \begin{array}{c} s \\ b \end{array} \right)_{L}},
\end{displaymath}

\par
The weak interaction carriers $A^{i}_{\alpha}$, which are responsible
for the weak transitions between different quark families are connected
with the $B, C, D$ bosons in the following manner:
$$
A^{1}_{\alpha} \to B^{\pm}_{\alpha} , A^{2}_{\alpha} \to C^{\pm}_{\alpha} ,
A^{3}_{\alpha} \to D^{\pm}_{\alpha}.
\eqno(16)
$$
\par
Using the data from [1] and equation (14) we have obtained the following
masses for $B^\pm, C^\pm, D^\pm, E^\pm$ bosons:
\par
$
m_{B^{\pm}} \cong  169.5 \div 171.8\hbox{ GeV.},\\
$
\par
$
m_{C^{\pm}} \cong  345.2 \div 448.4\hbox{ GeV.},\hspace{5cm}(17)\\
$
\par
$
m_{D^{\pm}} \cong 958.8 \div 1794$ GeV.,\\
\par
$
m_{E^{\pm}} \cong  4170 \div 4230\hbox{ GeV..}\\
$
Now consider some development of our model.
\par
a. It is clear that the masses of quarks and $B, C, D, E$ bosons
can be introduced using the Higgs's mechanism.
Here arises a question about correspondence of the physical picture given
by Higgs's mechanism to the real physical picture of quarks and vector bosons.
In the Higgs's mechanism the quarks and vector bosons get their masses
through their interactions with Higgs's bosons [11] (in an analogy with the
mechanism of superconductivity), i.e. in the presence of Higgs's fields the
quarks and vector bosons are massive. It is clear that free quarks and
vector bosons (in reality, we have free quarks and vector bosons) must be
massless. Then we see that Higgs's mechanism is
for introducing masses in the theory without singularity
(i.e. without straight violation gauge invariance), but not a mechanism of
masses generation.
\par
On the other side, the standard weak interaction
cannot generate masses for its $\gamma_5$ invariance.
\par
Then
the following question arises: how are masses of these particles
generated?
\par
It is obvious that these quarks and bosons must have a structure i.e.,
they consist of subparticles which take part in some interactions which
generate masses. So, we see that it goes in an analogy with the
strong interactions, where the fundamental interaction  is the
chromodynamics and the hadrons consist of the quarks. It is clear that
if the quarks and massive bosons consist  of subparticles, then in our
approach (the Model of Dynamical Analogy of the Cabibbo-Kobayashi-Maskawa
matrices) the problem of singularity does not appear since at small
distances interact subparticles but not quarks and massive bosons.
And then the problem of singularity must be solved in the theory of
subparticle interactions in full analogy with the strong interactions
theory. It is obvious that in the framework of our model it is not
needed to use GIM mechanism [10] to cancel the singularity.
\par
b. Let us have $K^\pm$ which is produced in strong interactions and
we want to consider its decay. Since $K$ meson includes $s$ quark, then
when we take into account the weak interaction, we must use the Cabibbo
matrix [2] mixing $s, d$ quarks:
$$
\begin{array}{c}
d' =  d cos \theta + s sin \theta  \\
s'  =  -d sin \theta  + s cos \theta
\end{array}
\eqno(18)
$$
i.e., $s$ quark transforms in superpositions of $s, d$ quarks
$$
s  \rightarrow  -d sin \theta  + s cos \theta
\eqno(19)
$$
\par
The matrix element of $K$ meson decay [7] is proportional to
$sin \theta$, i.e., we take into account only the $sin\theta$ part
from expression (19) and then the term proportional to
$cos\theta$ is remained. It means that  only the part proportional
to $sin\theta$ decays. However, from the current experiments we
know that $K$ mesons decay fully. It can happen only if $K$
mesons decay through massive bosons $B$ but not $W$ boson and
the $sin\theta$ term of Cabibbo matrix. Then the mass of this massive
boson $B$ must be determined through the following expression:
$$
m_b^2 \cong \frac{m_W^2}{sin \theta}
\eqno(20)
$$
We see that this massive boson is like $B$ boson which appears in
the above considered model of dynamical analogy of
Kabibbo-Kobayashi-Maskawa matrices [6].
\par
Let us pass to a more detailed consideration of the virtual oscillation
case since it is of a real interest (i.e. we compute nondiagonal term of
the mass matrix).

\section {The $\pi \stackrel{B}\leftrightarrow K$ Meson Transitions
in Diagram Approach in the
Model of Dynamical Analogy of Kabibbo-Kobayashi-Maskawa Matrices }

When one takes into account $d, s$ quark mixings and $B$ exchange,
the diagram for $\pi \stackrel{B} \longrightarrow K$ transitions has
the form \\

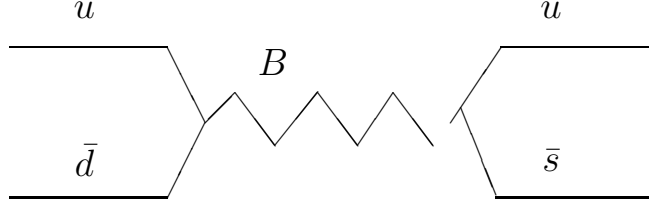
\unitlength=1.00mm
\special{em:linewidth 0.4pt}
\linethickness{0.4pt}
\begin{picture}(113.00,50.00)
\put(29.00,39.00){\line(1,0){20.00}}
\put(49.00,39.00){\line(1,-2){5.00}}
\put(54.00,29.00){\line(-1,-2){5.00}}
\put(49.00,19.00){\line(-1,0){20.00}}
\put(29.00,19.00){\line(0,0){0.00}}
\put(29.00,19.00){\line(0,0){0.00}}
\put(29.00,19.00){\line(0,0){0.00}}
\put(29.00,19.00){\line(0,0){0.00}}
\put(54.00,29.00){\line(1,1){4.00}}
\put(58.00,33.00){\line(3,-4){5.33}}
\put(63.33,26.00){\line(4,5){5.67}}
\put(69.00,33.00){\line(3,-4){5.33}}
\put(74.33,26.00){\line(2,3){4.67}}
\put(79.00,33.00){\line(3,-4){5.33}}
\put(84.33,26.00){\line(5,6){2.33}}
\put(86.67,29.00){\line(2,3){6.67}}
\put(93.33,39.00){\line(1,0){19.67}}
\put(88.00,31.00){\line(2,-5){4.67}}
\put(92.67,19.00){\line(1,0){20.33}}
\put(28.00,39.00){\line(1,0){1.00}}
\put(28.00,19.00){\line(1,0){2.00}}
\put(38.00,44.00){\makebox(0,0)[cc]{$u$}}
\put(38.00,24.00){\makebox(0,0)[cc]{$\bar d$}}
\put(63.00,37.00){\makebox(0,0)[cc]{$B$}}
\put(100.00,44.00){\makebox(0,0)[cc]{$u$}}
\put(100.00,24.00){\makebox(0,0)[cc]{$\bar s$}}
\end{picture}

\par
It is clear that at $d, s$ mixings the transition of $\pi$ meson mass
shell does not take place, i.e. $K$ meson produced from $\pi$ meson
remains on the mass shell of $\pi$ meson.
\par
The amplitude of this process has the following form (we use Feynman
rules):
$$
M(\pi \rightarrow K) = {G_B} [\bar d \gamma_\mu
(1 - \gamma_5) u][\bar s \gamma^\mu (1 - \gamma_5) u] ,
$$
or
$$
M(\pi \rightarrow K) = {G_B} [\bar d Q_\mu u][\bar s Q^\mu u] ,
\eqno(21)
$$
where $G_B$ is Fermi of $B$ boson constant which is connected with
Fermi constant $G_W$ of $W$ by the following relation
$$
G_B = G_F sin\theta, \qquad {G_F\over \sqrt{2}} = {g^2\over 8m_W^2} ,
$$
 and
$Q_\mu = \gamma_\mu (1 - \gamma_5)$  .
\par
The mass Lagrangian $L$ for this diagram in the framework of standard
approach is [7]
$$
L = M(\pi \rightarrow K) .
\eqno(22)
$$
Then the mass differences in squared form which response for
$\pi \rightarrow K$ and $K \rightarrow \pi$ transitions is
$$
m^2_1 - m^2_2 = <\pi\mid L \mid K> + <K \mid L \mid \pi>
\eqno(23)
$$
(we suppose that $K$ meson is on the mass shell of $\pi$ meson).
Therefore
$$
m^2_1 - m^2_2 \simeq 2 m_\pi \Delta m_{1 2}
\eqno(24)
$$
or
$$
\Delta m_{1 2} = \frac{1}{2m_\pi} [<\pi\mid L \mid K> + <K \mid L \mid \pi>]
\eqno(25)
$$
Now we compute mass difference. For this goal we use the following
expressions:
$$
\begin{array}{c}
<0 \mid \bar d Q_\mu u \mid \pi> \phi_\pi f_\pi p_\mu, \\
<0 \mid \bar s Q^\mu u \mid K> = \phi_K f_K p^\mu ,
\end {array}
\eqno(26)
$$
where $\phi_\pi, \phi_K $ , $f_\pi, f_K$, correspondingly, are the
wave functions and the constant decays of  $\pi$ and $K$ mesons,
$p_\mu$ is momentum of $\pi$ meson.
\par
It is necessary to remark that the following relation for
constant decays on mass shells will be:
$$
f_\pi (m_\pi) = f_K (m_\pi),
\eqno(27)
$$
Then from equation (23) using equations (26), (27) we obtain the
following expression:
$$
\Delta m^2 =
m^2_1 - m^2_2 = f^2_\pi m_\pi^2 {G_B} .
\eqno(28)
$$
or (see Eq. (4))
$$
m_{\pi K} = \Delta m_{1 2} = f^2_\pi m_\pi {G_B} .
\eqno(29)
$$

\section{Probability of $\pi \stackrel{B}\longleftrightarrow K$
Virtual Oscillations with Account of $\pi$ Decays}

\par
If at $t = 0$ we have the flow $N(\pi, 0)$ of $\pi$ mesons, then at
$t \ne 0$ this flow will decrease since $\pi$ mesons decay and then
we have the following flow $N(\pi, t)$ of $\pi$ mesons:
$$
N(\pi, t) = exp(- \frac{t}{\tau_0}) N(\pi, 0)  ,
\eqno(30)
$$
where $\tau_0 = \tau'_0 \frac{E_\pi}{m_\pi}$.
\par
The expression for the flow $N(\pi \rightarrow K, t)$, i.e. probability of $\pi$ to $K$ meson transitions at time $t$, has the following form:
$$
N(\pi \rightarrow K, t) = N(\pi, t) P(\pi \rightarrow K, L)
\eqno(31)
$$
where
$$
P(\pi \rightarrow K, L) =
sin^2 \left[\pi \frac{L}{L_{osc}} \right ] ,
$$
$$
L_{osc} = \frac{2.48 p_{\pi}(MeV)}{\mid m_1^2 - m_2^2 \mid (eV^2)} m .
$$
and
$$
m^2_1 - m^2_2 = f^2_\pi m_\pi^2 {G_B} .
$$
\par
The expression for probability of $\pi \rightarrow K$ oscillations
$P(\pi \rightarrow K, t)$, in the approach where
the phase volume is taken into account, has the following form [4,5]:
$$
N(\pi \rightarrow K, t) =
N(\pi, t) sin^2 \left[\frac{\pi t}{\tau(\pi \stackrel{B} \longrightarrow K)}
\right ] =
$$
$$
= N(\pi, 0) exp(-\frac{t}{\tau_0})sin^2\left[\frac{\pi t}{\tau_0}
\frac{\frac{m_W^4}{m_B^4}}{(\frac{m_\mu}
{m_u + m_{\bar d}})^2}\right] .
\eqno(32)
$$
\par
Probability of $\pi \rightarrow K$ real oscillations
$P(\pi \rightarrow K, t)$ in the case of real oscillations is described
by the following expression (see Eq. (8)) [4,5]:
$$
P(\pi \rightarrow K, t) =  sin^2 2\beta
sin^2 \left[\pi t \frac{m_K^2}{2 p} \right ] ,
$$
where
$$
sin^2 2\beta \cong \frac{4m^2_{\pi K}}{(m_\pi - m_K)^2}  \cong 0 .
$$
\par
The kinematics of $K$ meson production processes in quasielastic
processes is given in work [4].

\section{Conclusion}

The elements of the theory of vacuum oscillations and the model of
dynamical expansion of the theory of weak interactions works at the
tree level, i.e. the model of dynamical analogy of
Cabibbo-Kobayashi- Maskawa matrices and its further development, were given.
It was shown that the quarks and massive vector bosons must be structural
and these structural particles (subparticles) must interact to
generate quark and vector boson masses.
In this case  the problem of singularity cancellations does not arise
in this model.
It was also shown
that, for self consistency of the theory, the weak decays of $K$ mesons
must go through massive vector boson $B$ but not $W$ boson.
\par
In the framework of this model the probability of $\pi \leftrightarrow K$
transitions (oscillations) in the diagram approach was computed.
These transitions are virtual since masses of $\pi$ and $K$ mesons
differ considerably.  These transitions (oscillations) can be registered
through $K$ decays after transitions of virtual $K$ mesons to their own
mass shell by using their quasielastic strong interactions.

\begin{center}
REFERENCES  \\
\end{center}
\par
\medskip
\noindent
1. Review of Particle Prop., Phys. Rev. 1992, D45, N. 11.
\par
\noindent
2. Cabibbo N., Phys. Rev. Lett., 1963, 10, p.531.
\par
Kobayashi M. and Maskawa K., Prog. Theor. Phys., 1973, 49,
\par
p.652.
\par
\noindent
3. Beshtoev Kh.M., JINR Commun. E2-93-167, Dubna, 1993;
\par
   JINR Commun. E2-95-326, Dubna, 1995;
\par
   Chinese Journ. of Phys., 1996, 34, p.979.
\par
\noindent
4. Beshtoev Kh.M., JINR Commun. E2-98-387, Dubna, 1998.
\par
\noindent
5. Beshtoev Kh.M., JINR Commun. E2-99-137, Dubna, 1999.
\par
\noindent
6. Beshtoev Kh.M., JINR Commun. E2-94-293, Dubna, 1994;
\par
   Turkish Journ. of Physics, 1996, 20, p.1245;
\par
   JINR Commun. E2-95-535, Dubna, 1995;
\par
   JINR Commun. P2-96-450, Dubna, 1996.
\par
   JINR Commun. E2-97-210, Dubna, 1997.
\par
\noindent
7. Okun L.B., Leptons and Quarks, M. Nauka, 1990.
\par
\noindent
8. Blatt J.M., Weisscopf V.F., The Theory of Nuclear Reactions,
\par
ONR T. R. 42.
\par
\noindent
9. Maiani L., Proc.Int. Symp.  on  Lepton-Photon Inter., Hamburg,
\par
   DESY, p.867.
\par
\noindent
10. Glashow S., Iliopoulos J. and Maiani L., Phys. Rev., 1970,
\par
   D2, p.1285.
\par
\noindent
11. Higgs P.W., Phys. Rev. Lett. 1964, 12, p.1132;
\par
Phys. Rev., 1966, 145, p.1156.

\end{document}